\documentclass[preprints,article,accept,moreauthors,pdftex]{Definitions/mdpi} 

\firstpage{1} 
\pubvolume{10}
\issuenum{10}
\articlenumber{1952}
\pubyear{2020}
\copyrightyear{2020}
\history{Received: 6 September 2020; Accepted: 27 September 2020; Published: 30 September 2020}
\updates{yes} 

\usepackage{upgreek}

\Title{Integrated 
	Source of Path-Entangled Photon Pairs with Efficient Pump Self-Rejection}

\Author{Pablo de la Hoz $^{1,}$*\orcidA{},  Anton Sakovich $^{2}$, Alexander Mikhalychev $^{2}$\orcidB{}, {Matthew Thornton} $^1$\orcidC{}, {Natalia Korolkova} $^1$\orcidD{}  and Dmitri Mogilevtsev $^2$\orcidE{} }

\AuthorNames{Pablo de la Hoz, Anton Sakovich, Alexander Mikhalychev, Matthew Thornton, Natalia Korolkova, Dmitri Mogilevtsev }

\address{%
$^{1}$ \quad School of Physics and Astronomy, University of St Andrews,
North Haugh, St Andrews KY16 9SS, UK; mt45@st-andrews.ac.uk (M.T.); nvk@st-andrews.ac.uk (N.K.)\\
$^{2}$ \quad B. I. Stepanov Institute of Physics, National Academy of Sciences of Belarus, Nezavisimosti Ave. 68-2, 220072 Minsk,  Belarus; sakovich.2.718281828459045@gmail.com (A.S.); mikhalychev@gmail.com (A.M.); d.mogilevtsev@ifanbel.bas-net.by (D.M.)}

\corres{Correspondence: pdlhi1@st-andrews.ac.uk}

\abstract{We present a theoretical proposal for an integrated four-wave mixing source of narrow-band path-entangled photon pairs with efficient spatial pump self-rejection. The scheme is based on correlated loss in a system of waveguides in Kerr nonlinear media. We calculate that this setup gives the possibility for upwards of 100 dB pump rejection, without additional filtering. The effect is reached by driving the symmetric collective mode that is strongly attenuated by an engineered dissipation, while~photon pairs are born in the antisymmetric mode. A  similar set-up can additionally be realized for the generation of two-photon NOON states, also with pump self-rejection. We discuss the implementation of the scheme by means of the coherent diffusive photonics, and demostrate its feasibility in both glass (such as fused silica-glass and IG2) and planar semiconductor waveguide structures in indium phosphide (InP) and in silicon.}


\begin{document}


\section{Introduction}

The generation of photon pairs is a staple tool of modern quantum technologies. Twin photons have found far-reaching applications in a wide range of fields, from~quantum communications to imaging, metrology, and LIDARs~\cite{lloyd,padjet1,deg}. One of the established methods for producing photon pairs is the spontaneous four-wave mixing  (SFWM) process in Kerr-nonlinear structures~\cite{agarwal,fiorentino,caspiani}. This method is very promising, with~the perspective to create integrated sources of photon pairs that are compatible with the other photonic blocks necessary, for~example, for~quantum processors or quantum key distribution systems~\cite{knill,diam}.
Such a process can be realized in integrated waveguiding structures (for~example, in~silicon or indium phospide (InP) platforms), which render them very suitable for building quantum photonic circuits~\cite{sukhoruk,silicon,kruse,oser}.  

One of the main problems in implementing SFWM  for photon pair generation  is pump rejection, especially with CW pumping. Because, in the SFWM process, two photons are converted into signal and idler photons, and~all four are of close frequencies, achieving large pump rejection can be challenging and it requires quite exquisite filtering~\cite{caspiani}. This is often accomplished via filtering setups far larger in size that the nonlinear device producing photon pairs. Recently, there has been great interest in producing on-chip filters that reject the pump~\cite{sukhoruk,oser}. 
Generally, the~majority of these filtering schemes are based on very precise frequency filtering allowing for transmission bands less than 1~nm wide and achieving  more than 100 dB transmission of pass-band to stop-band contrast~\cite{sukhoruk,oser,piek,diego}. Only~recently, a theoretical proposal using photonic crystals has appeared suggesting to exploit spatial features, i.e.,~suppress coupling of the output mode to the pump by symmetry considerations~\cite{savona}. Similar spatial filtering ideas were also suggested for the integrated source of photon pairs based on the three-wave mixing by the spontaneous parametric down-conversion (SPDC)~\cite{spat,spat1}.

\begin{figure}[H]
\begin{center}
\includegraphics[width=0.65\linewidth]{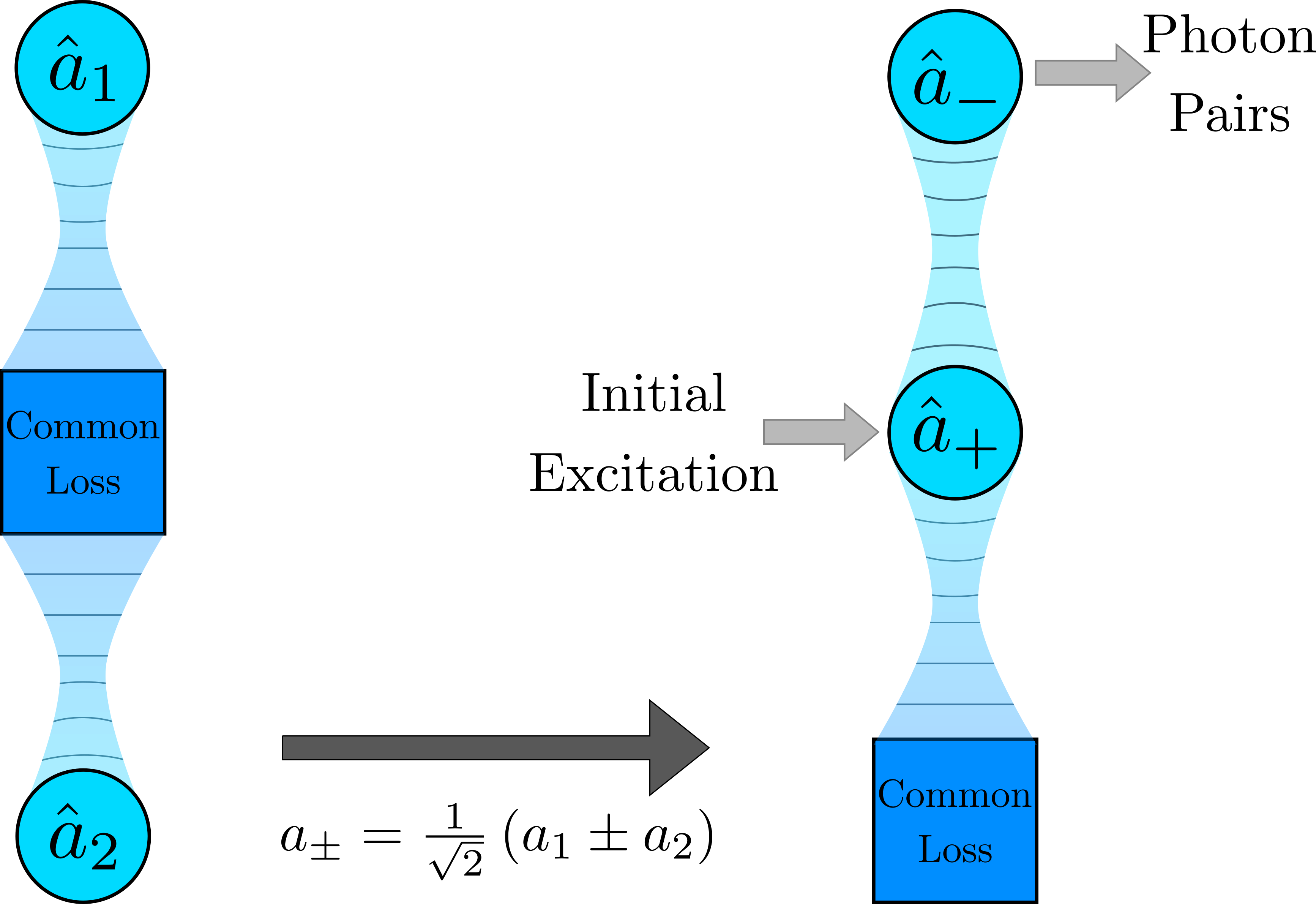}
\end{center}
\caption{The scheme of the photon-pair generator with in-built pump rejection on the basis of two single-mode waveguides coupled to the common loss~reservoir.} 
\label{fig1}
\end{figure}

In this work, we suggest a novel and simple way for realizing efficient on-chip pump rejection by the SFWM process in a waveguide structure with engineered loss. In~essence, this structure consists of two waveguides coupled only dissipatively through a common reservoir, as in Figure~\ref{fig1}. Coupling to the common reservoir defines a superposition mode subject to a strong engineered loss. If~one only pumps this modal superposition, photon pairs are born in the orthogonal mode and they travel through the waveguides, whereas the pump exponentially decays along the waveguide. Our scheme allows for the pump to be filtered out, even if the pump, signal, and idler are of the same~frequency. 

Such a ``built-in'' filtering can be further enhanced by exploiting the symmetry properties of the modes (as it was done in recent works with SPDC~\cite{spat,spat1}): the output from both waveguides can be interfered on a beamsplitter, in~order  to filter out the possible remnants of the symmetric mode and put both photons in the same spatial~mode. 

Furthermore, our pump-rejecting scheme can be modified in order to produce a pair of photons in two collective modes (i.e., in~four spatial modes), which can then be used to produce two-photon NOON states~\cite{mitchell, dowling}. These are multi-photon entangled states corresponding to the superposition of $N = 2$ photons in the first mode with zero photons in the second mode, and~vice~versa.

Note, that systems containing Kerr-type nonlinearities have already been considered as a source of entangled states~\cite{kowalewska, olsen, kalaga}. Systems of two coupled waveguides with Kerr nonlinear medium used in one or both channels have been studied in the context of co- and contra-directional couplers (for review, see~\cite{perina00}). Quantum statistics and dynamics of Kerr nonlinear couplers with linear and nonlinear coupling, including regimes of varying linear coupling, have been attracting interest for more than three decades now  and shown to lead to the whole range of different non-trivial quantum effects, such as collapses and revivals of oscillations, sub-Poissonian photon statistics in single and in compound modes~\cite{korolkova97, korolkova97b}, as~well as  higher-order nonclassicalities: quantum entanglement, squeezing, and~antibunching~\cite{thapliyal}.  Already, the possibility to achieve novel regimes that emulate the dynamics of complex many-body systems has been investigated (see, e.g.,~\cite{mogilevtsev97}).
Nowadays, a whole new active field of research has emerged, topological photonics~\cite{meany}, where also our dissipatively coupled systems may contribute.  The~principle difference of the systems that is presented in this paper to the nonlinear directional couplers is the use of the engineered non-linear loss as the main mechanism determining the device functionality~\cite{mog2010, natcom, prap}.

\section{Scheme}
\label{sec:scheme}

Let us introduce the simplest model to describe the \textit{modus operandi} of our generator with in-built pump~rejection. 

The basic scheme that we suggest for the pair generation is just two identical, self-Kerr nonlinear single-mode waveguides solely coupled by common loss in a symmetric way (Figure~\ref{fig1}). The mode dynamics in this device can be described by the following master equation for the density matrix $\rho$:
\begin{eqnarray}
\frac{d}{dt}\rho=-i\frac{U}{2} [(a_1^{\dagger})^2a_1^2+(a_2^{\dagger})^2a_2^2,\rho]+
\frac{1}{2}\Gamma L(a_1+a_2)\rho+\gamma (L(a_1)+L(a_2))\rho,
\label{basic}    
\end{eqnarray}
where $a_j$, $a_j^{\dagger}$ are bosonic annihilation and creation operators of $j$-th mode, $U$ is the nonlinearity, $\gamma$ and $\Gamma$ are individual and collective loss rates, and~the dissipator is $L(a_j)\rho=a_j\rho a_j^{\dagger}-\frac{1}{2}\rho a_j^{\dagger}a_j-\frac{1}{2}a_j^{\dagger}a_j\rho $. 

After transforming to the basis $a_{\pm}=\frac{1}{\sqrt{2}}(a_1\pm a_2)$, Equation~(\ref{basic}) becomes 
\begin{eqnarray}
\frac{d}{dt}{\bar\rho}=-i[H+V,{\bar\rho}]+
(\Gamma+\gamma) L(a_+){\bar\rho}+\gamma L(a_-){\bar\rho},\label{basic2}
\end{eqnarray}
where the Kerr interaction Hamiltonian is
\begin{eqnarray}
H=\frac{U}{4}\Bigl(n_+^2+n_-^2+4n_+n_--n_+-n_-  \Bigr),    
\label{h}
\end{eqnarray}
and the two-photon exchange Hamiltonian is
\begin{eqnarray}
 V=\frac{U}{4}((a_+^{\dagger})^2a_-^2+h.c.).   
 \label{v}
\end{eqnarray}
the operators $n_{\pm}=a_{\pm}^{\dagger}a_{\pm}$ are  photon-number operators for the superposition~modes. 

The scheme of two self-Kerr nonlinear modes coupled by the common loss was already considered in a number of works~\cite{valer,mog2010,natcom,prap}. However, it was mostly considered as a way to engineer two-photon loss for generating non-classical states from the initial classical input. Collective loss was assumed to be strong, and~the symmetric mode $a_+$ was usually adiabatically~eliminated. 

Here, we exploit a different and somewhat counter-intuitive strategy. Let us initially excite only the symmetric mode  $a_+$. In~realistic waveguiding structures, Kerr nonlinearity is small. Hence, if,~initially, the symmetric mode is in a coherent state and~ $U\langle n_+\rangle \ll \Gamma$, both the Kerr nonlinearity and interaction with the antisymmetric mode will hardly affect the symmetric mode. Its state will remain coherent and uncorrelated with the state of the antisymmetric~mode.  

Under this assumption, after~averaging over the states of the symmetric mode, Equation~(\ref{basic2}) becomes the following equation for the single-mode density matrix $\varrho$:
\begin{eqnarray}
\frac{d}{dt}{\varrho}\approx -i[H_-+V_-(t),{\varrho}]+\gamma L(a_-){\varrho},
\label{basic3}    
\end{eqnarray}
where the driving-independent part is $H_-=\frac{U}{4}(n_-^2-n_-)$.  The~driving term reads
\begin{eqnarray}
V_-(t)=U|\alpha_+(t)|^2n_- + \frac{U}{4}\left(\alpha^2(t)(a_-^{\dagger})^2+h.c.\right),  
\label{v2}
\end{eqnarray}
with
\begin{eqnarray}
  \alpha_+(t)\approx \alpha_+(0)\exp\{-\frac{1}{2}(\Gamma+\gamma)t\},   
    \label{reject}
\end{eqnarray}
where $\alpha_+(0)$ is the initial amplitude of the symmetric~mode. 

For simplicity sake, let us first assume that the waveguide loss is negligibly small, $\gamma T\ll 1$, where $T$ is the total propagation time, and~the driving is not very strong,  ${U}|\alpha_+(0)|^2/\Gamma\ll 1$. For~the probability of the two-photon generation, one derives the following result from Equations~(\ref{basic3}) and (\ref{v2})
\begin{eqnarray}
P_2(t)\propto \frac{U^2|\alpha_+(0)|^4}{\Gamma^2}(1-\exp\{-\Gamma t\})^2.
\label{prob2}
\end{eqnarray}

\noindent
Equations~(\ref{reject}) and (\ref{prob2}) describe the action of the "built-in" pump rejection. The~pump exponentially decays, whereas the probability of the pair creation approaches its maximal value for the propagation time when the pump is almost completely rejected. One~can arbitrarily enhance pump~rejection by~simply increasing the propagation time. 

Of course, the~unavoidable presence of waveguide loss limits the possible extension of the waveguide, as~it leads to the destruction of the generated photon pairs. However, our scheme has an additional intrinsic mechanism of pump rejection: spatial symmetry. A~common 50/50 beamsplitter at the outcome of our device allows for the remnant of the  driving field (which is in the symmetric mode) to be filtered out. A~similar spatial-filtering mechanism was recently suggested for pump rejection for the SPDC-based waveguide source of photon pairs~\cite{spat,spat1}.

\section{Results and~Discussions}
\vspace{-6pt}
\subsection{Operational~Regime}

Let us clarify the conditions of operation for our pump-rejecting pair generator.
We want to have, at the output of the device, an average number of photon pairs that is much higher than the numbers of surviving pump photons and single photons appearing after destruction of the pairs by unavoidable realistic linear loss. 
The condition of our scheme functioning at some time $T$ can be given in the following way
\begin{eqnarray}
P_2(T)\gg \frac{1}{2}P_1(T),\frac{1}{2}|\alpha_+(T)|^2,
\label{condition}
\end{eqnarray}
where $P_1(T)$ is the probability of single-photon generation at the time $T$. 

When one takes into account the linear loss $\gamma$ and considers its rate, $\gamma<<\Gamma$, as~being  much lower than engineered loss, $\Gamma$, for~a weak pump  $(U|\alpha_+(0)|^2/\Gamma\ll 1)$, the~probability of the two photon generation can be derived from Equations~(\ref{basic3})--(\ref{reject}), as:
\begin{eqnarray}
P_2(t)\approx\frac{1}{8}\frac{U^2|\alpha_+(0)|^4}{\Gamma^2}\exp\{-2\gamma t\}(1-\exp\{-\Gamma t\})^2.
\label{p2_loss}
\end{eqnarray}

\begin{figure}[H]
\begin{center}
\scalebox{1}[1]{\includegraphics[width=\textwidth]{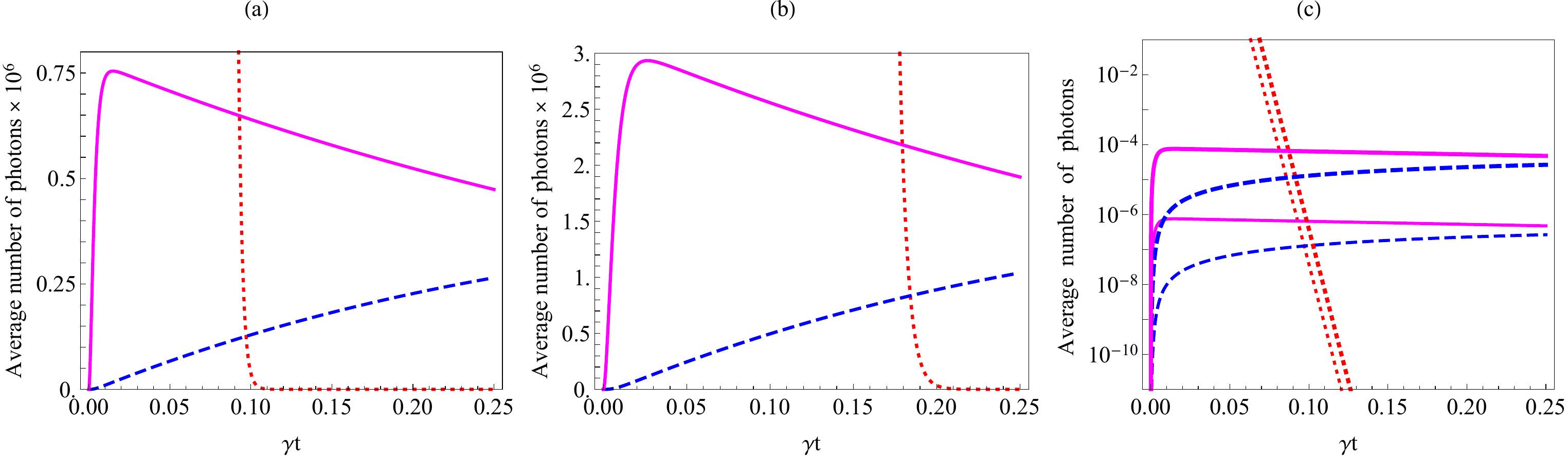}}
\end{center}
\caption{Average number of twin-photons (solid lines), single-photons (dashed lines), and pump photons (dotted lines) as given by Equations~(\ref{reject}), (\ref{p2_loss}) and (\ref{p1_loss}). (\textbf{a},\textbf{b}) correspond to $\Gamma=400\gamma$ and $\Gamma=200\gamma$ respectively and the initial average number of photons of the pump $|\alpha_+(0)|^2=10^{10}$. Figure~(\textbf{c}) corresponds to $|\alpha_+(0)|^2=10^{10}$ (thin lines) and $|\alpha_+(0)|^2=10^{11}$ (thick lines) and $\Gamma=400\gamma$. For~all of the figures, the squared nonlinearity is $U^2=10^{-20}\gamma^2$.} 
\label{fig2}
\end{figure}

 Equations~(\ref{reject}) and (\ref{p2_loss}) show that, for the condition (\ref{condition}) to be fulfilled in the presence of linear loss, one needs the interaction to take place over a time interval that is larger than the time that maximizes the probability of pair generation.
Indeed, from~Equation~\eqref{p2_loss}, an~estimate for the time of the maximal probability of obtaining a photon pair is given by:
\begin{eqnarray}
t_{\rm max}\approx-\frac{1}{\Gamma}\ln \left(\frac{\gamma}{\Gamma }\right).
\label{tmax}
\end{eqnarray}
one can see that the optimal time (\ref{tmax}) does not yield a large pump rejection. For~pump intensity, the~degree of rejection is just the ratio of  loss rates, $x=\frac{\gamma}{\Gamma }$. However, it is easy to see that $n$-time increase in the interaction time over $t_{max}$ drastically suppresses the pump (as $x^n$), but~leads to relatively small decrease of two-photon generation probability. Figure~\ref{fig2} illustrates this situation for a rather large ($\sim$$10^{10}$) initial number of pump photons.  Notice that, for the illustrated case, about 150 dB suppression of the pump takes~place. 

The \textls[-15]{probability of single-photons as result of the pair decay can be derived from Equations~(\ref{basic3})--(\ref{reject}), as}
\begin{eqnarray}
\label{p1_loss}
P_1(t)\approx \frac{1}{4}\frac{U^2|\alpha_+(0)|^4}{\Gamma^2}\exp\{-\gamma t\}\left(1-\exp\{-\gamma t\}-\frac{2\gamma}{\Gamma}(1-\exp\{-\Gamma t\})+
\frac{\gamma}{2\Gamma}(1-\exp\{-2\Gamma t\})\right)
\end{eqnarray}
so, one can see from Equations~(\ref{reject}), (\ref{p2_loss}) and (\ref{p1_loss}) that to fulfill the condition (\ref{condition}), the~interaction time $T$ should satisfy
\begin{eqnarray}
\frac{1}{\gamma}\gg T >\frac{1}{\Gamma}\ln{\left(\frac{4\Gamma^2}{ U^2|\alpha_+(0)|^2\delta}\right)},
\label{condition2}
\end{eqnarray}
where the parameter $\delta=|\alpha_+(T)|^2/2P_2(T)\ll 1$ defines the acceptable level of the pump~rejection.

The condition (\ref{condition2}) imposes a limit on the relation between the "natural" linear loss $\gamma$ and the engineered one, $\Gamma$, for~our scheme to be functional and to provide for generation of photon pairs with low admixture of single-photons. Generally,  $\Gamma\gg\gamma$ is required. For~the example that is shown in  Figure~\ref{fig2}a,b,  the~engineered loss rate should be at least two orders of magnitude larger than the natural loss rate to give a probability of photon pair generation much exceeding that of~single-photons. 

As we demonstrate later on, such a large engineered loss is completely feasible and it can be easily achieved while using different material platforms (for example, waveguides in fused silica or planar waveguides in InP). Of~course, an~increase in the engineered loss rate will lower the pair generation probability. However, this can be compensated by an increase in pump intensity. As~it follows from the condition (\ref{condition2}), even a large increase in pump intensity still does not lead to a significant increase of the interaction time that is required for the pump rejection. Figure~\ref{fig2}c illustrates this situation: a~tenfold increase in pump intensity leads to a hundredfold increase of  $P_{1,2}$, but~only few percent change in the required interaction time (see Equations~(\ref{p2_loss}) and (\ref{p1_loss})). 

\unskip


\subsection{Quantum Perturbation~Theory}

In this Section, we confirm that  theresults of the previous section, obtained in the approximation of a coherent pump, remain valid, even beyond this approximation, i.e.,~when both modes are treated using quantum perturbation theory. These results continue to hold provided that the weak pump approximation remains valid,  $\lambda=U|\alpha_+(0)|^2/\Gamma\ll 1$. Using a perturbation theory for the operators, in~ Appendix~\ref{apA} the following expressions are derived for the average numbers of photons in each collective mode:
\begin{align}
 \nonumber
 \left\langle n_+(t)\right\rangle&=|\alpha_+(0)|^2e^{-(\Gamma+\gamma) t}+ \mathcal{O}(\lambda^2), \\
  \label{2mode}
 \left\langle n_-(t)\right\rangle&=\frac{U^2|\alpha_+(0)|^4}{2\Gamma(2\Gamma+\gamma)}
 \frac{e^{-2(\Gamma+\gamma) t}}{\Gamma+\gamma}\times
 \left\{ \eta(t)\Gamma+\gamma \zeta(t)\right\}+ \mathcal{O}(\lambda^3),
\end{align}
with functions $\eta(t)=\text{e}^{(2\Gamma+\gamma)t}-2 \text{e}^{\Gamma t}+1$ and $\zeta(t)=1-\text{e}^{\Gamma t}$.

The average number of pump photons $\left\langle n_+(t)\right\rangle$ agrees with Equation~\eqref{reject} that was obtained by the semiclassical approximation. Moreover, a~higher-order correction that is given by Equation~(\ref{higherorder}) shows that, even when the average number of the pump photons is comparable with the average number of generated photons, the~deviation from the semiclassical formula Equation~\eqref{reject}  is only~small. 

We also check that the result (\ref{2mode}) for the average number of photons in the antisymmetric mode, $\left\langle n_-(t)\right\rangle$, for~$\gamma\ll\Gamma$ corresponds to the sum $2P_2(t)+P_1(t)$, as given by Equations~(\ref{p2_loss}) and (\ref{p1_loss}). 

Thus, we deduce that, in the limit of the weak pump, $U|\alpha_+(0)|^2/\Gamma\ll 1$, semiclassical pump approximation gives results that are very close with the quantum analysis up to the very low numbers of~photons.

\subsection{Realizations}

\begin{figure}[H]
\begin{center}
\includegraphics[width=0.5\linewidth]{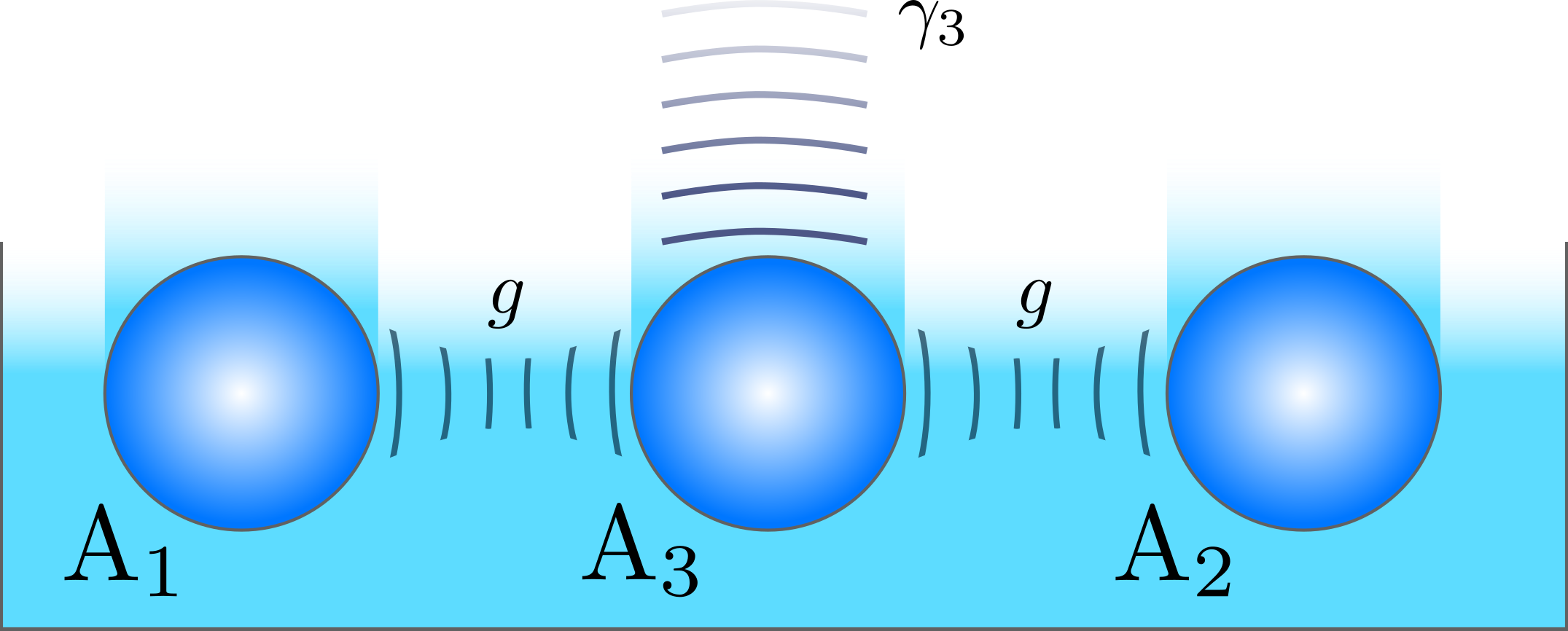}
\end{center}
\caption{ An example of three-waveguide realization of the basic Scheme (\ref{basic}). The~waveguide $A_3$ is subjected to engineered loss with the rate $\gamma_3$. Both waveguides $A_1$ and $A_2$ are unitary coupled to the waveguide $A_3$; the coupling constant is $g$.} 
\label{fig3}
\end{figure}

The simplest way to realize the collective loss that is described in the Scheme  (\ref{basic}) is to couple two single-mode waveguides to a third lossy waveguide (see Figure~\ref{fig3}). Subsequently, for~strong loss in the middle waveguide, one can adiabatically exclude the third mode ($A_3$ in Figure~\ref{fig3}) and arrive at the master Equation~(\ref{basic}). Under~symmetric coherent excitation of both waveguides $A_{1,2}$, the~regime of the pair generation can be realized. This scheme has been suggested for the realization of different kinds of nonlinear loss~\cite{valer,mog2010}, and~for dissipative beamsplitting/equalization~\cite{natcom}.  The~adiabatic elimination was discussed in detail in the recent work~\cite{prap}. Basically, if~the side single-mode waveguides are coupled to the central one with the coupling rates $g$, and~the central waveguide is subjected to the linear loss with the rate $\gamma_3$, the~collective loss rate is
\begin{equation}
\Gamma\approx 8\frac{g^2}{\gamma_3}.
\label{clr}
\end{equation}
Equation~(\ref{clr}) shows that ratios of the engineered and "natural" loss rates given by condition~(\ref{condition2}) are easily feasible in three-waveguide structures that are depicted in Figure~\ref{fig3}. For~example, in~the recent works~\cite{natcom,prap}, an all-glass scheme with laser-inscribed single-mode waveguides was realized with the strong linear loss in the third waveguide induced by a long "tail" of coupled waveguides.  In~such "tailed" glass structures, one can routinely have $g$ of about 200--300 m$^{-1}$ in the infrared-visible wavelength regions, and~a rate $\gamma_3$ of about four times their value, allowing for $\Gamma$ of about 400--600 m$^{-1}$.  Even for  highly nonlinear glass, such as chalcogenide glass IG2, with~losses about 12 m$^{-1}$ at 1 $\upmu$m wavelength, the~condition (\ref{condition2}) can be easily satisfied. For~common silica glass, propagation loss can be less than  2 m$^{-1}$ in the infrared and optical wavelength regions~\cite{nasu,grev}. Even better ratios are possible while using drawing techniques for waveguide structures (which is commonly used for producing photonic crystal fibers~\cite{pcf}). There, the propagation loss can be almost as low as for the conventional index-guiding optical single-mode fibers, and~be lower than $10^{-4}$ m$^{-1}$~\cite{niels}.

\begin{figure}[H]
\begin{center}
\includegraphics[width=0.7\linewidth]{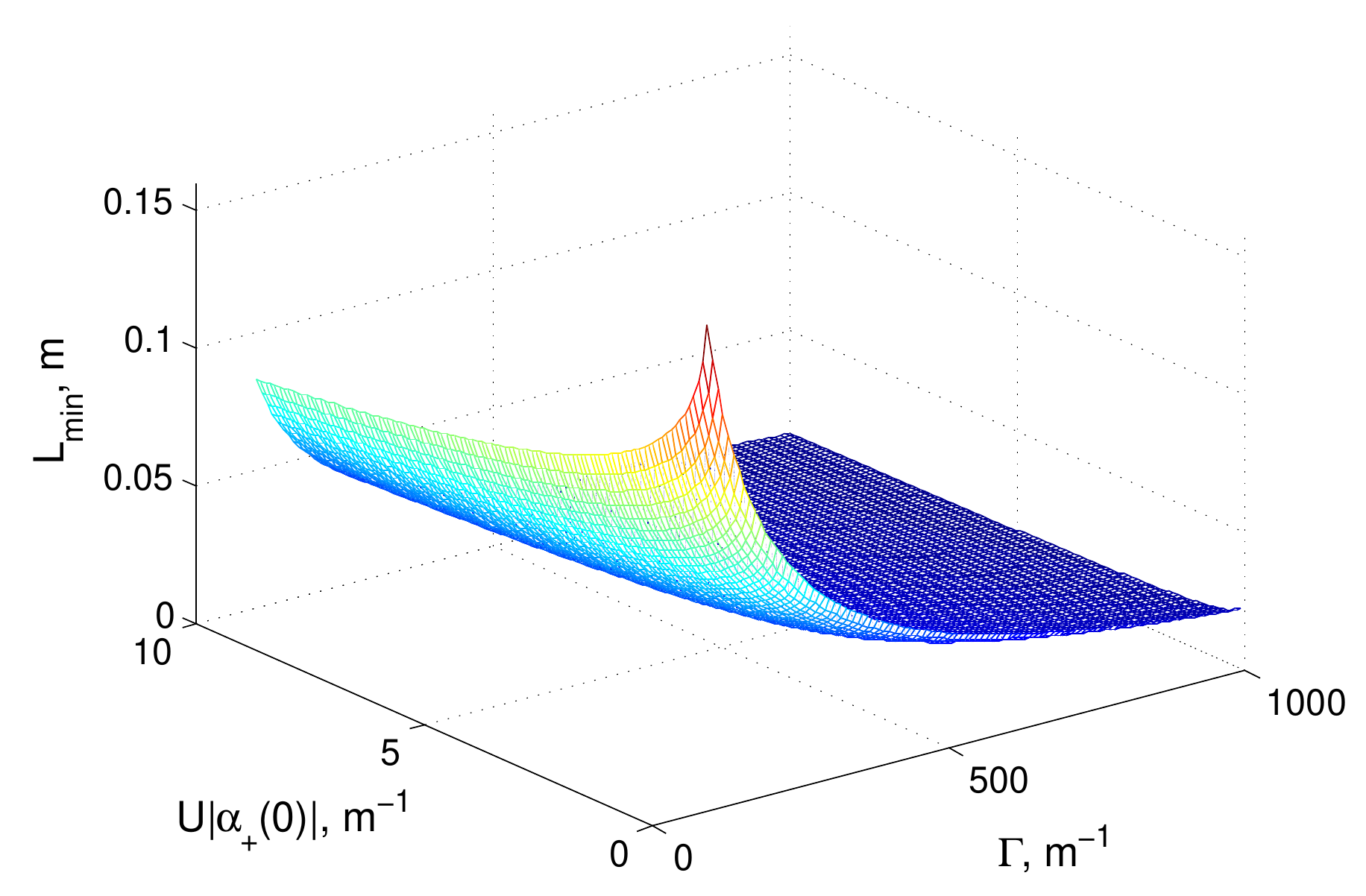}
\end{center}
\caption{A minimal waveguide length, $L_{min}$, providing for pump rejection with $\delta=0.1$ for typical parameters of the glass waveguides and weak~pump.} 
\label{fig4}
\end{figure}

Obviously, all-glass structures might face size problems when integrated into optical circuits.  For~example, while the width of the structure for  800 nm wavelength will be less than 50 $\upmu$m, the~height of the "tail" (which should be of more that 10 waveguides~\cite{natcom,prap}) might be 0.5 mm and more. More serious is the problem with the necessary propagation length. 
Figure~\ref{fig4} shows the minimal interaction length, $L_{min}$, providing for pump rejection with $\delta=0.1$, as~given by the left-hand side of the condition~(\ref{condition2}) and for the typical system parameters that are discussed in this Section.  One can see that, for quite a wide range of parameters, the~minimal length that is required to reduce the average number of residual pump photons much below the number of generated photons, is of about few cm. This limits the possibilities of~integration.

However,  one can strongly reduce the required device length by combining the pump self-rejection and spatial rejection scheme: the residual pump can be rejected by 50/50 beamsplitter at the output of the device. Indeed, increasing the coupling strength in order to provide an order of magnitude larger $\Gamma$ than those that are typical in all-glass structures, one can achieve more than 50 dB pump self-rejection with a structure of just few millimeter length. This approach seems to be useful with planar ridge waveguides built on such common platforms as silicon~\cite{sil} or indium phosphide-(InP)~\cite{inp}-based planar semiconductor structures. These can provide for a possibility to have much higher coupling (and, correspondingly, $\Gamma$) than in glass, at~the price of having much larger propagation loss (for example, of~about 50--100 m$^{-1}$ for InP waveguides~\cite{inp} or silicon waweguides~\cite{sil}).

Of course, an increase in $\Gamma$ leads to a decrease in the pair generation rate. However, the~larger nonlinearity and tighter mode confinement more than compensate for it. Indeed, for~the device length larger than $\Gamma^{-1}$, from~Equation~(\ref{p2_loss}), the~pair generation rate is given by~\cite{prap}
\begin{eqnarray}
    R_2\approx \left(\frac{2\pi n_2}{\lambda S}P_w\right)^2\frac{c}{8\Gamma} \eta,
    \label{2rate}
\end{eqnarray}
where $P_w$ is the input power, $c$ is the speed of light, $\lambda$ is the wavelength, $S$ is the waveguide modal area, $n_2$ is  the nonlinear refractive index, and $\eta=e^{-2\gamma T}$ is extinction coefficient describing $P_2(T)$ reduction by single-photon~loss. 

For fused silica waveguides with typical values of $n_2\approx 2\times10^{-20}~$m$^2 $W$^{-1}$, for~$g = 300~$m$^{-1}$ and negligibly small linear loss, at~$\lambda=800~\mathrm{nm}$ for a single-mode waveguide with typical modal area of $5\times10^{-11} \upmu$m$^2$, about 180 W input power is needed in order to reach of about 1 KHz pair generation rate. This~power value can be lowered by implementing weaker waveguide coupling for low-loss long fiber-like structures. However, as~Equation~(\ref{2rate}) shows, it is much more advantageous to take a material with higher nonlinearity and tighter field localization, than~to aim for lower $\Gamma$ and longer~structures. 

For IG2-glass waveguides with the same modal area and $n_2$ of about two orders of magnitude larger than for fused silica, it is sufficient to have less than $200~\upmu$W for reaching the same generation rate. With~a single-mode silicon waveguides, one can have much tighter mode localization (for example, $ S\approx  5~\upmu$m$^2$ at 1550 nm wavelength~\cite{low}), and~much higher non-linearity. For~example, for~such a silicon waveguide with a large $\Gamma= 1000~$m$^{-1}$ at the wavelength of 1550 nm, just few $\upmu$W input power suffice for reaching the 1 KHz generation rate~\cite{sato}.  

Thus, the~highly nonlinear planar waveguide structure that is schematically shown in Figure~\ref{fig3}  is a prospective platform for realizing bright integrated generators of photon pairs with efficient pump~self-rejection.  

Further, our generator is robust with respect to the Raman scattering noise that commonly arises in photon pair generators implementing SFWM~\cite{agr}. Such noise manifests as  uncorrelated photons in waveguides and they can be the dominant source of noise in SFWM pair-generating devices~\cite{raman}.  However, in~our devices, the Raman scattering produces uncorrelated photons in the symmetric mode, which is subject to  strong engineered~loss. 

Finally, we notice that imperfections in the waveguide structure leading to the asymmetric coupling between waveguides might negatively affect the discussed scheme. Asymmetry will lead to the generation of uncorrelated single photons in the antisymmetric mode~\cite{mog2010}. However, as~long as the discrepancy in coupling constants is not large in comparison with their values, the~effect is negligible. More precisely, if~the first waveguide couples to the dissipative waveguide with the rate $g+\delta>0$, and~the second waveguide couples to the dissipative waveguide with the rate $g-\delta>0$, the~effect of asymmetry will be negligible for $g\gg 4|\delta|$~\cite{mog2010}. In~practice, it means that the deviations in distance between waveguides should be much lower than this distance, which is perfectly feasible~\cite{natcom}.


\subsection{Extension to NOON~States}

\begin{figure}[H]
\begin{center}
\includegraphics[width=0.48\linewidth]{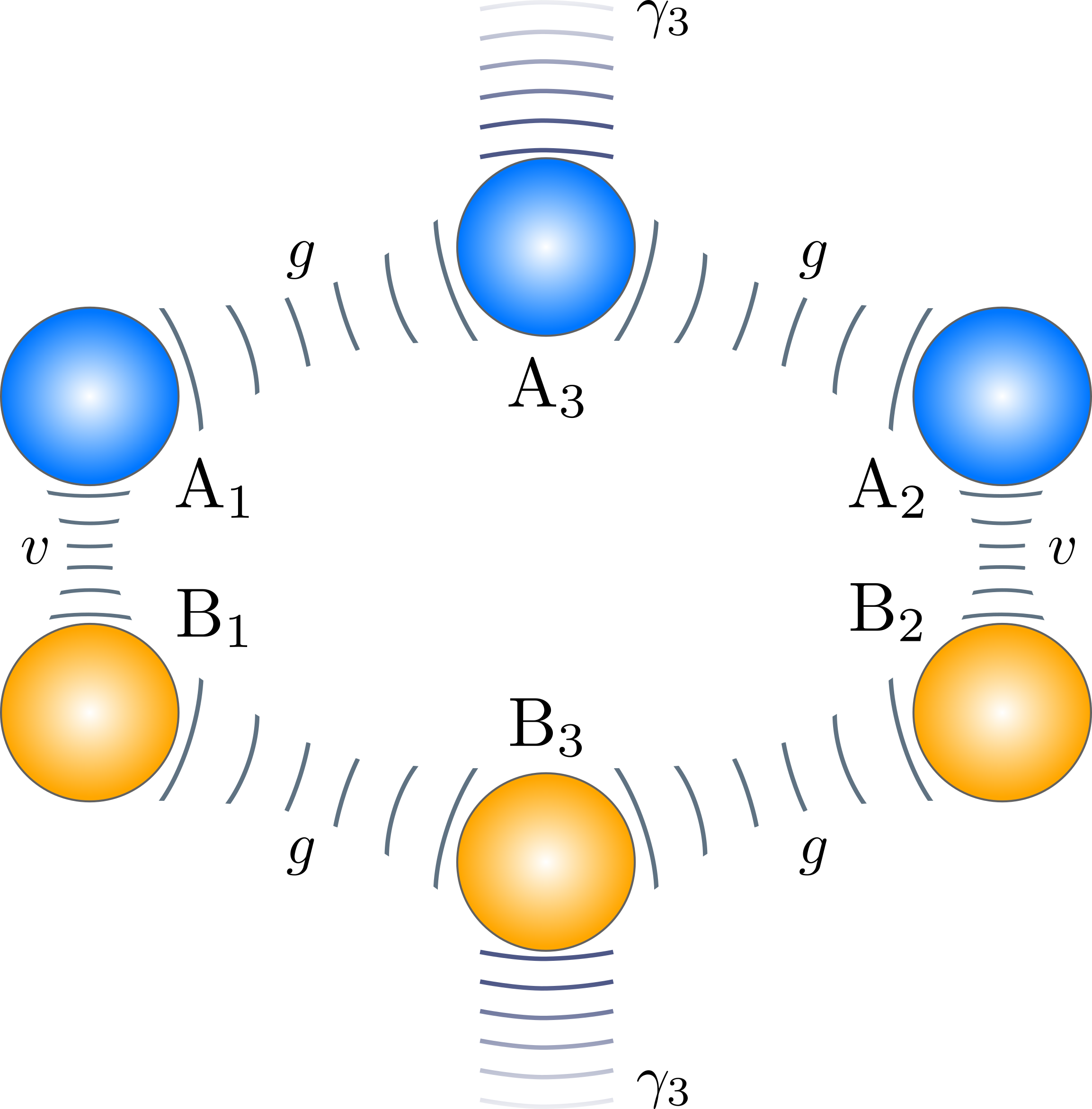}
\end{center}
\caption{ An example of six-waveguide realization of the NOON state generation scheme. The~depicted structure shows two mirrored three-waveguide structures of Figure~\ref{fig3} with waveguides labeled as $A_j$ for the upper, and~$B_j$ for the lower structures. The~waveguides $A_3$ and $B_3$ are subject to strong engeneered loss. The~waveguide $A_1$ is unitary coupled with the waveguide $B_1$, the~waveguide $A_2$ is unitary coupled with the waveguide $B_2$; the coupling constant is $v$. Other parameters are shown in Figure~\ref{fig3}.} 
\label{fig5}
\end{figure}

\begin{figure}[H]
\begin{center}
\includegraphics[width=0.48\linewidth]{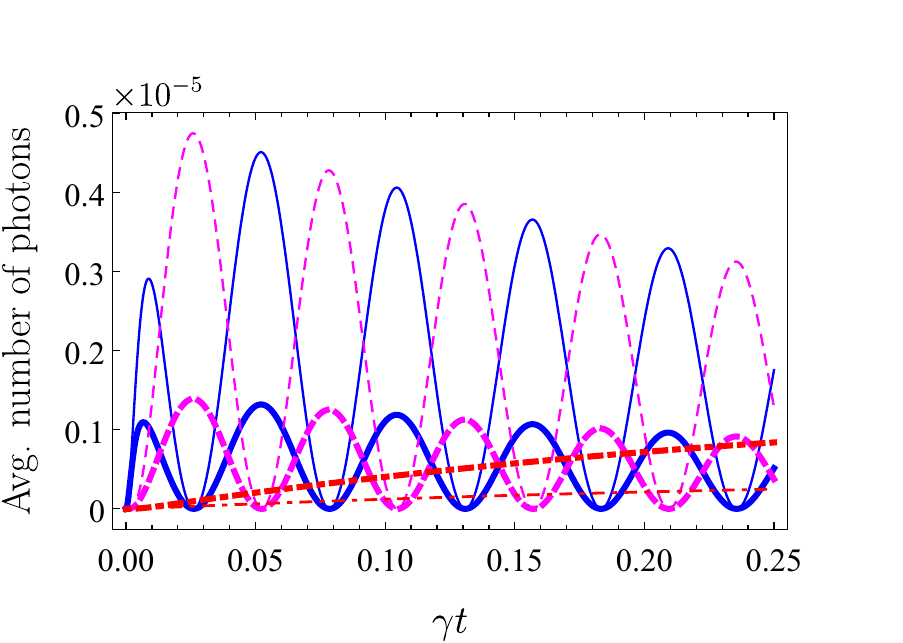}
\end{center}
\caption{Average number of photons corresponding to the generation of two photons in one of the antisymmetric modes (blue solid lines) and of two photons being simultaneously in both antisymmetric modes (magenta dashed lines) according to Equation~\eqref{noon11} (they are, respectively, $2\times P_{2,0}$ and $P_{1,1}$). Red~dashed-dotted lines show the probability of finding a single photon only in one of the antisymmetric modes, due to photon loss from one of the previous configurations.  The coupling constant has been chosen to be $v=30 \gamma$. Thick lines correspond to $\Gamma= 400 \gamma$, thin lines correspond to $\Gamma= 200\gamma$. Other parameters are the same as in Figure~\ref{fig2}.  } 
\label{fig6}
\end{figure}

The described principle of the pump self-rejection can also be applied for more complicated setups than just the two dissipatively symmetrically 
coupled waveguides considered so far in this work. Any~coherent diffusive photonic circuit working by collective loss (see Ref.~\cite{natcom} for several examples of such devices) can be implemented for the pump self-rejection scheme, as it is described above. As~an example, let us consider a scheme that allows for us to simultaneously produce single-photon states in different spatial modes.  Two-photon NOON states may then be generated by interfering these photons. The~setup allowing this, Figure~\ref{fig5}, consists of two mirror-imaged two-mode devices considered before. The~state-carrying waveguides are unitarily coupled in the usual way, while the modes $A_1$, $B_1$, and~$A_2$, $B_2$ are additionally coupled with rate $v$. This single-photon exchange process is described by the standard Hamiltonian , $W=\hbar v(a_1^{\dagger}b_1+a_2^{\dagger}b_2+h.c.)$, where $v$ is the coupling rate, the~operators $a_j$ describe state-carrying modes of the upper half of the device, and~operators $b_j$ describe modes of the lower half of the device (the whole scheme is considered in detail in Appendix~\ref{apB}). It is easy to see that, under Hamiltonian $W$, only the modes with the same symmetry are coupled. If~one puts into all four  state-carrying modes  ($A_{1,2}$, $B_{1,2}$) coherent states with the same amplitude (say, $\alpha$), then in the limits of low ``natural'' loss, low pump intensity and large interaction time $T\gg 1/{\Gamma}$, the~photon-number probabilities read
\begin{eqnarray}
\nonumber
P_{1,1}\propto \frac{U^2|\alpha(0)|^4}{\Gamma^2}(\sin\{2vT\})^2, \\ 
P_{2,0}\propto \frac{U^2|\alpha(0)|^4}{\Gamma^2}(\cos\{2vT\})^2
\label{noon11}
\end{eqnarray}
probability $P_{1, 1}(T)$ denotes the probability to simultaneously generate two photons, one in each of the antisymmetric superposition modes, while $P_{2, 0}(T)$ denotes the probability to generate two photons in just one of the antisymmetric modes. Equation~(\ref{noon11}) show that, for any designed device parameters, one~can always choose the coupling constant $v$ (for example, just by adjusting the distance between the upper and lower parts of the device), as~to have a photon in each antisymmetric mode at the output, and~produce two-photon NOON state by interfering these photons. Figure~\ref{fig6} visualizes the~relevant probabilities, where we include the effects of linear~loss. 

\unskip
\section{Conclusions}

We have suggested and discussed a scheme of pump self-rejection for the generation of photon pairs and two-photon path-entangled states by the four-wave mixing process in the system of Kerr nonlinear waveguides. The~cornerstone of our scheme is the strong collective loss acting on the waveguides  pairs. This loss strongly affects the symmetric collective modes, but~leaves the antisymmetric mode intact.  By~driving the symmetric collective mode, we obtain photon pairs in the antisymmetric mode. For~a sufficiently long device, the~pump is completely filtered out, but photon loss and the appearance of uncorrelated single~photons limit the length. 

We have analyzed device design and performance for the realistic waveguide structures in both glass and for such platforms as indium phosphide (InP) and silicon based planar semiconductor structures. We have demonstrated that the available nonlinearities, waveguide coupling rates, and~``natural'' linear loss will allow for the fabrication of functional, high-quality photon-pair generators with pump self-rejection. In~addition, one can combine self-rejection with the spatial rejection: it is possible to filter out the remnant of the pump just by interfering the output modes and in order to separate the antisymmetric~mode. 

We have also demonstrated the way to use our pump self-rejection for more complicated photonic circuits for the generation of entangled photons. Duplicating our two-waveguide set-up, it is possible to simultaneously produce two photons in two different group of waveguides, and~generate NOON~states. 


\vspace{6pt} 



\authorcontributions{Conceptualization, D.M. and N.K.; methodology, D.M. and N.K.; software, A.S. and M.T.; validation, A.S. and P.d.l.H.; formal analysis, P.d.l.H., A.M. and D.M.; investigation, P.d.l.H., A.S., A.M., N.K., M.T. and D.M.; resources, D.M. and N.K.; writing--original draft preparation, P.d.l.H. and D.M.; writing--review and editing, M.T.; visualization, M.T.; supervision, N.K. and D.M.; project administration, N.K.; funding acquisition, A.M., D.M., N.K. All authors have read and agreed to the
published version of the~manuscript.}

\funding{This research was supported by the EU Flagship on Quantum Technologies, project PhoG (820365). D.M., A.S. and A.M. also acknowledge support from the National Academy of Sciences of Belarus program~``Convergence''. The authors gratefully acknowledge the support from the Scottish Universities Physics Alliance (SUPA) and the Engineering and Physical Sciences Research Council (EPSRC). }

\conflictsofinterest{The authors declare no conflict of~interest. } 



\appendixtitles{yes} 
\appendix
\section{Heisenberg-Picture Operator Perturbation Expansion for the Photon~Number}\label{apA}
{To check the validity of the predictions obtained by the simplest scheme, and~in order to investigate the dynamical regime in which the results start to deviate from those obtained from the basic assumptions made in Section~\ref{sec:scheme}, the~two-mode description of our photon source is explored.  To~do this, we consider the two mode master equation Equation~\eqref{basic2}, which is expressed in terms of the collective modes.}

{One could solve directly the master equation to obtain the time evolution of the density matrix and subsequently use the properties of the trace to obtain the observable quantities. However, an~important simplification of the problem is achieved if we consider the evolution of the operators in the Heisenberg picture~\cite{dezfouli}. Thus, we start with the adjoint master equation:
\begin{eqnarray}
    \frac{d}{dt}{A(t)}=i[H+V,A(t)]+(\Gamma+\gamma) L^\dagger(a_+){A(t)}
    +\gamma L^\dagger(a_-){A(t)},
    \label{adj}
\end{eqnarray}
for any operator $A$. Since the Hamiltonian of Equation~\eqref{h} is diagonal in the Fock state basis, it does not manifest itself when we calculate the probability of photon pair generation, nor when we calculate the expected value for the photon number in each collective mode. This way we can focus our attention in the interaction Hamiltonian Equation~\eqref{v} describing the conversion of two-photons from the symmetric mode into the antisymmetric one, and~vice~versa.}

{The presence of the non-linear Hamiltonian precludes an exact solution of Equation~\eqref{adj}. However, in~the conditions previously mentioned $\lambda=U|\alpha_+(0)|^2/\Gamma$ is a small parameter, and~therefore we can obtain approximate solutions for the evolution of the operator $A$ in the form of a power series in the small parameter, where we require the interaction Hamiltonian $V=\mathcal{O}(\lambda)$.}

{We consider a perturbation expansion of the photon number operator:
\begin{eqnarray}
\begin{aligned}
    \left(a_{\pm}^{\dagger} a_{\pm}\right)(t)=\left(a_{\pm}^{\dagger} a_{\pm}\right)^{(0)}(t)+\left(a_{\pm}^{\dagger} a_{\pm}\right)^{(1)}(t)+\left(a_{\pm}^{\dagger} a_{\pm}\right)^{(2)}(t)+\cdots
\end{aligned}
\end{eqnarray}
where the $j$-th term of the expansion is proportional to $\lambda^j$, i.e.,~$\left(a_{\pm}^{\dagger}a_{\pm}\right)^{(j)}=\mathcal{O}(\lambda^j)$
}

{
We can solve Equation~\eqref{adj} for $a_{\pm}^{\dagger}a_{\pm}$    by
solving an equivalent set of equations which are
found by identifying the terms of the same order in
$\lambda$. These are equations on  $\left(a_{\pm}^{\dagger}a_{\pm}\right)^{(j)}$ that can be solved once we know the solution for $\left(a_{\pm}^{\dagger}a_{\pm}\right)^{(j-1)}$. In~particular, we have:
\begingroup\makeatletter\def\f@size{9}\check@mathfonts
\def\maketag@@@#1{\hbox{\m@th\normalsize \normalfont#1}}%
\begin{align}
\frac{d}{d t}\left(a_{\pm}^{\dagger} a_{\pm}\right)^{(0)}(t)&=\Gamma_-\mathcal{L}^{\dagger}\left(a_{-}\right)\left(a_{\pm}^{\dagger} a_{n}\right)^{(0)}(t)
+ \Gamma_+ \mathcal{L}^{\dagger}\left(a_{+}\right)\left(a_{\pm}^{\dagger} a_{\pm}\right)^{(0)}(t),
\label{zero}\\
\frac{d}{d t}\left(a_{\pm}^{\dagger} a_{\pm}\right)^{(j)}(t)&=i \left[V,  \left(a_{\pm}^{\dagger} a_{\pm}\right)^{(j-1)}(t)\right]+\Gamma_-\mathcal{L}^{\dagger}\left(a_{-}\right)\left(a_{\pm}^{\dagger} a_{\pm}\right)^{(j)}(t)+ \Gamma_+ \mathcal{L}^{\dagger}\left(a_{+}\right)\left(a_{\pm}^{\dagger} a_{\pm}\right)^{(j)}(t) 
\label{jth}
\end{align}
\endgroup
with $\Gamma_+=\Gamma+\gamma$ and $\Gamma_-=\gamma$.
The zero-th order Equation \eqref{zero} can be readily solved as it does not contain the Hamiltonian. As~the initial condition, we require that the Heisenberg and Schr\"odinger operators coincide at $t=0$.
We understand in what follows that the operators without explicit time dependence are Schr\"odinger operators. We reach the zero-th order  solution:
\begin{eqnarray}
\left(a_{\pm}^{\dagger} a_{\pm}\right)^{(0)}(t)& = a_{\pm}^{\dagger} a_{\pm} e^{-\Gamma_\pm t}
\end{eqnarray}
Photons in the symmetric and antisymmetric modes decay exponentially
at rates $\Gamma + \gamma$ and $\gamma$ respectively. The~knowledge of the zero-th order solution allows us to solve for the first order:
\begin{eqnarray}
\begin{aligned}
       \left(a_{\pm}^{\dagger} a_{\pm}\right)^{(1)}(t)=\frac{i U e^{- (\Gamma +2 \gamma)t} \left(e^{ \Gamma_{\mp}t}-1\right)}{2 \Gamma_{\mp}}\left(a_\pm^\dagger a_\pm^\dagger a_\mp a_\mp - a_\mp^\dagger a_\mp^\dagger a_\pm a_\pm\right)
       \label{1order}
\end{aligned}    
\end{eqnarray}
Similarly, we obtain the second order solution:
\begingroup\makeatletter\def\f@size{9}\check@mathfonts
\def\maketag@@@#1{\hbox{\m@th\normalsize \normalfont#1}}%
\begin{eqnarray}
\begin{gathered}
    \left(a_{\pm}^{\dagger} a_{\pm}\right)^{(2)}(t)=f_{1\pm}^{(2)}(t)(a_\pm^\dagger)^2a_\pm^2+f_{2\pm}^{(2)}(t)(a_\mp^\dagger)^2 a_\mp^2+f_{3\pm}^{(2)}(a_\pm^\dagger)^2a_\pm^2(a_\mp^\dagger) a_\mp + f_{4\pm}^{(2)}(t)(a_\pm^\dagger) a_\pm(a_\mp^\dagger)^2 a_\mp^2
    \label{2order}
\end{gathered}    
\end{eqnarray}
\endgroup
where $f_{i\pm}^{(2)}(t)$ with $i=1,\dots 4$ are functions of time that determine the dynamical evolution at second order. The~explicit expression for these functions can be obtained after substitution of the expressions Equations~\eqref{1order} and \eqref{2order} in Equation~\eqref{jth}. The~result is a differential equation for each of these functions that we obtain by identifying the coefficients of the corresponding operators. After~integration, we obtain:
\begin{align}
\label{corrections3}
    f_{1\pm}^{(2)}(t)=&\pm\frac{U^2 e^{- (\Gamma+2 \gamma  )t}}{2 \Gamma  \Gamma _{\mp} \left(\pm \Gamma +\Gamma _{\mp}\right)}\left[\Gamma _{\mp} \left(1-e^{\mp \Gamma t  }\right)\pm\Gamma  \left(1-e^{\Gamma _{\mp}t }\right)\right],\\
     f_{2\pm}^{(2)}(t)=& \mp \frac{U^2 e^{-( \Gamma+2 \gamma )t } }{2 \Gamma  \Gamma _{\mp} \left( 3 \Gamma _{\mp}-2 \gamma -\Gamma\right)}\left[(2 \gamma +\Gamma ) \left(1-e^{\Gamma _{\mp}t }\right) +\Gamma _{\mp} \left(2 e^{\Gamma _{\mp}t }+e^{\pm \Gamma t }-3\right)\right],\\
    f_{3\pm}^{(2)}(t)=&-\frac{U^2 e^{-(2\Gamma_\pm+\Gamma_{\mp})t} }{\gamma (\Gamma +\gamma) (\Gamma +2 \gamma)}\left[\Gamma_{\mp}+\Gamma_{\pm} e^{ (\Gamma +2 \gamma)t}-(\Gamma +2 \gamma) e^{ \Gamma_{\pm}t}\right],\\
    f_{4\pm}^{(2)}(t)=&\frac{U^2  e^{- (2\Gamma_{\mp}+\Gamma_{\pm} )t}}{2 \Gamma_{\mp}^2}\left(e^{\gamma_{\mp} t}-1\right)^2
\end{align}}
the photons-number statistics that characterize our source of photon pairs can be obtained from the analytic expressions we have now derived. We assume that the symmetric mode is initially excited in a coherent state. The~average values in the Heisenberg picture are obtained as $\langle A (t) \rangle= \text{Tr}[A(t)\rho_0]$ with $\rho_0=|\alpha_+(0)\rangle\langle\alpha_+(0)|\otimes|0\rangle\langle 0|$. For~the symmetric mode:
\begin{align}
    \left\langle n_+(t)\right\rangle=&\langle n_+^{(0)}(t) \rangle+ \langle n_+^{(2)}(t)\rangle + \mathcal{O}(\lambda^3),\\
   \langle n_+^{(0)}(t)\rangle=&|\alpha_+(0)|^2e^{-(\Gamma+\gamma) t},\\
   \langle n_+^{(2)}(t)\rangle=&-\frac{U^2|\alpha_+(0)|^4}{2 \Gamma  \gamma (\Gamma +\gamma)} e^{-2  (\Gamma +\gamma)t} \left(\gamma+\Gamma  e^{ (\Gamma +\gamma)t}-(\Gamma +\gamma) e^{\Gamma  t}\right),
   \label{higherorder}
\end{align}
for the antisymmetric mode:
\begin{align}
    \left\langle n_-(t)\right\rangle&= \langle n_-^{(2)}(t)\rangle + \mathcal{O}(\lambda^3),\\
   \langle n_-^{(2)}(t)\rangle&=\frac{U^2|\alpha_+(0)|^4}{2\Gamma(2\Gamma+\gamma)}\frac{e^{-2(\Gamma+\gamma) t}}{\Gamma+\gamma}\left( \eta(t)\Gamma+\gamma \zeta(t)\right),
\end{align}
with functions 
$\eta(t)=\text{e}^{(2\Gamma+\gamma)t}-2 \text{e}^{\Gamma t}+1$
and $\zeta(t)=1-\text{e}^{\Gamma t}$.

\section{NOON State~Generator }\label{apB}
The pair-generator scheme considered until now in this work is produces pairs of photons in the same (antisymmetric) mode. However, a~simple modification of the scheme allows us to create photons in two different superposition modes, and~thus to generate two-photon NOON states. This~modification is two three-waveguide structures, as~of Figure~\ref{fig3}, with~unitary coupling between the state-carrying modes depicted in Figure~\ref{fig5}. After~elimination of the modes $a_3$ and $b_3$, this system is described by the following master equation
\begin{eqnarray}
\nonumber
\frac{d}{dt}\rho=\sum\limits_{x,j}\left(-i\frac{U}{2} [(x_j^{\dagger})^2x_j^2,\rho]+
\gamma L(x_j)\rho\right)-
iv[a_1^{\dagger}b_1+a_2^{\dagger}b_2+b_1^{\dagger}a_1+b_2^{\dagger}a_2,\rho]+\label{basicnoon} \\
\frac{1}{2}\Gamma \left(L(a_1+a_2)+L(b_1+b_2)\right)\rho,  
\end{eqnarray}
where $x=a,b$, and~$a_j$ denote the modes of upper half of the device, and~$b_j$ denote modes of the lower half of the device. The~constant $v$ describes the coupling between the waveguides of the both~halves. 

After transforming to the basis $x_{\pm}=\frac{1}{\sqrt{2}}(x_1\pm x_2)$, Equation~(\ref{basicnoon}) becomes
\begingroup\makeatletter\def\f@size{9}\check@mathfonts
\def\maketag@@@#1{\hbox{\m@th\normalsize \normalfont#1}}%
\begin{eqnarray}
\frac{d}{dt}{\bar\rho}=-\sum\limits_{x=a,b}i[H_x+V_x,{\bar\rho}]-
iv[(a_+^{\dagger}b_++a_-^{\dagger}b_-)+h.c.),\rho]+
\sum\limits_{x=a,b}\left((\Gamma+\gamma) L(x_+)+\gamma L(x_-)\right){\bar\rho},\label{basicnoon2} 
 \end{eqnarray}
 \endgroup
where the Kerr interaction Hamiltonian is
\begin{eqnarray}
H=\frac{U}{4}\Bigl(n_{x+}^2+n_{x-}^2+4n_{x+}n_{x-}-n_{x+}-n_{x-}  \Bigr),    
\label{hnoon}
\end{eqnarray}
and the two-photon exchange Hamiltonian is
\begin{eqnarray}
 V_x=\frac{U}{4}((x_+^{\dagger})^2x_-^2+h.c.)   
 \label{vnoon}
\end{eqnarray}
the operators $n_{x\pm}=x_{\pm}^{\dagger}x_{\pm}$ are  photon-number operators for the superposition modes. Notice that the photon exchange between symmetric and antisymmetric modes in the scheme described by Equations~(\ref{basicnoon2})--(\ref{vnoon}) occurs only through the two-photon exchange (\ref{vnoon}). 

 In the interaction picture with respect to the unitary coupling Hamiltonian $W=v[a_-^{\dagger}b_-+h.c.]$, considering the states of symmetric modes to be classical, after~averaging over the symmetric modes one obtains
\vspace{12pt}
\begin{eqnarray}
\frac{d}{dt}{\varrho}\approx -i[V(t),{\varrho}]+\gamma( L(a_-)+L(b_-)){\varrho},
\label{basicnoon3}    
\end{eqnarray}
where we have assumed both weak nonlinearity and weak pump. The~Hamiltonian:
\begin{eqnarray}
\nonumber
V(t)=\frac{U}{4}\sum\limits_{x=a,b}(n_{x-}^2+n_{x-}(4|\alpha_+(t)|^2-1)) + 
\frac{U}{4}\Bigl(\alpha^2(t)[(a_-^{\dagger})^2+(b_-^{\dagger})^2]\cos\{2vt\}+\label{noonv2}\\
2ia_-^{\dagger}b_-^{\dagger}\sin\{2vt\}+h.c.\Bigr) 
\end{eqnarray}
with
\begin{eqnarray}
  \alpha(t)\approx \alpha(0)\exp\{-\frac{1}{2}(\Gamma+\gamma)t-ivt\},   \label{noonreject}
\end{eqnarray}
where $\alpha(0)$ is the initial amplitude of both the symmetric modes $a_+$ and $b_{+}$. 

As it is with the simpler version of the scheme, Equation~(\ref{noonreject}) describes the exponential pump~rejection.  

To show how the NOON states generation functions, let us neglect an influence of the ``natural loss''. Then, for~large propagation time $T\gg 1/\Gamma$, and~for the unitary exchange rate much less than the engineered loss rate, $|v|\ll \Gamma$, after~returning to the original picture with respect to the unitary exchange operator $W$, one obtains  the following result for the probability of two photons being simultaneously in both antisymmetric superposition modes
\begin{eqnarray}
\nonumber
P_{1,1}\propto \frac{U^2|\alpha(0)|^4}{\Gamma^2}(\sin\{2vT\})^2 
\label{noon11a}
\end{eqnarray}

\noindent and the probability to have two photons in one of the antisymmetric 
modes
\begin{eqnarray}
\nonumber
P_{2,0}\propto \frac{U^2|\alpha(0)|^4}{\Gamma^2}(\cos\{2vT\})^2 
\label{noon2a}
\end{eqnarray}

\noindent interfering these two photons on the 50/50 beamsplitter, one gets the 2-photon NOON~state.

\reftitle{References}


\begin{thebibliography}{999}
\providecommand{\natexlab}[1]{#1}
\bibitem {lloyd} Pirandola, S.; Bardhan, B.R.; Gehring, T.; Weedbrook, C.; Lloyd, S. Advances in photonic quantum sensing. \emph{Nat. Phot.} \textbf{2018}, \emph{12}, 724. [\href{http://dx.doi.org/10.1038/s41566-018-0301-6}{CrossRef}]

\bibitem{padjet1} Moreau, P.-A.; Toninelli, E.; Gregory, T.; Padgett, M.J.  Imaging with quantum states of light. \emph{Nat. Rev. Phys.} \textbf{2019},\emph{ 1}, 367. [\href{http://dx.doi.org/10.1038/s42254-019-0056-0}{CrossRef}]

\bibitem{deg} Berchera, I.R.; Degiovanni, I.P. Quantum imaging with sub-Poissonian light: Challenges and perspectives in optical metrology. \emph{Metrologia} \textbf{2019},  \emph{56}, 024001. [\href{http://dx.doi.org/10.1088/1681-7575/aaf7b2}{CrossRef}]

\bibitem{agarwal} Agrawal, G.P.  \emph{Nonlinear Fiber Optics}, 3rd ed.; Academic Press: New York, NY, USA,  1995.

\bibitem{fiorentino} Fiorentino, M.; Voss, P.L.; Sharping, J.E.; Kumar, P.  All-fiber photon-pair source for quantum communications.  \emph{IEEE Photonics Technol. Lett.} \textbf{2002}, \emph{14},  983--985. [\href{http://dx.doi.org/10.1109/LPT.2002.1012406}{CrossRef}]

\bibitem{caspiani}  Caspani, L.; Xiong, C.; Eggleton, B.; Bajoni, D.; Liscidini, M.; Galli, M.; Morandotti, R.; Moss, D.J.   Integrated sources of photon quantum states based on nonlinear optics. \emph{Light Sci. Appl.} \textbf{2017}, \emph{6}, e17100. [\href{http://dx.doi.org/10.1038/lsa.2017.100}{CrossRef}] [\href{http://www.ncbi.nlm.nih.gov/pubmed/30167217}{PubMed}]


\bibitem{knill} Knill, E.; Laflamme, R.; Milburn, G.J. A scheme for efficient quantum computation with
linear optics.  \emph{Nature} \textbf{2001}, \emph{409}, 46. [\href{http://dx.doi.org/10.1038/35051009}{CrossRef}] [\href{http://www.ncbi.nlm.nih.gov/pubmed/11343107}{PubMed}]

\bibitem{diam} Diamanti, E.; Lo, H.-K.; Qi, B.; Yuan, Z. Practical challenges in quantum key
distribution. \emph{NPJ Quant. Inf.} \textbf{2016}, \emph{2}, 16025. [\href{http://dx.doi.org/10.1038/npjqi.2016.25}{CrossRef}]


\bibitem{sukhoruk} Solntsev, A.S.; Sukhorukov, A.A. Path-entangled photon sources on nonlinear chips. \emph{Rev. Phys.} \textbf{2017}, \emph{2}, 19--31. [\href{http://dx.doi.org/10.1016/j.revip.2016.11.003}{CrossRef}]


\bibitem{silicon} Silverstone, J.W.; Bonneau, D.; O’Brien, J.L.; Thompson, M.G. Silicon Quantum Photonics.  \emph{IEEE J. Sel. Top. Quantum Electron.} \textbf{2016}, \emph{22}, 390. [\href{http://dx.doi.org/10.1109/JSTQE.2016.2573218}{CrossRef}]

\bibitem{kruse} Kruse, R.; Sansoni, L.; Brauner, S.; Ricken, R.; Hamilton, C.S.; Jex, I.; Silberhorn, C.   Dual-path source engineering in integrated quantum optics. \emph{Phys. Rev. A} \textbf{2015}, \emph{92}, 053841. [\href{http://dx.doi.org/10.1103/PhysRevA.92.053841}{CrossRef}]

\bibitem{oser}  Oser, D.; Tanzilli, S.; Mazeas, F.; Alonso-Ramos, C.; Roux, X.L.; Sauder, G.; Hua, X.; Alibart, O.; Vivien, L.; Cassan, É.;~et~al. High-quality photonic entanglement out of a stand-alone silicon chip. \emph{NPJ Quantum Inf.} \textbf{2020}, \emph{6}, 31. [\href{http://dx.doi.org/10.1038/s41534-020-0263-7}{CrossRef}]

\bibitem{piek} Piekarek, M.; Bonneau, D.; Miki, S.; Yamashita, T.; Fujiwara, M.; Sasaki, M.; Terai, H.; Tanner, M.G.; Natarajan, C.M.; Hadfield, R.H.;~et~al.  High-extinction ratio integrated photonic filters for silicon quantum photonics. \emph{Opt. Lett.} \textbf{2017}, \emph{42}, 815. [\href{http://dx.doi.org/10.1364/OL.42.000815}{CrossRef}] [\href{http://www.ncbi.nlm.nih.gov/pubmed/28198872}{PubMed}]

\bibitem{diego} Pérez-Galacho, D.; Alonso-Ramos, C.; Mazeas, F.; Roux, X.L.; Oser, D.; Zhang, W.; Marris-Morini, D.; Labonte,~L.; Tanzilli, S.; Cassan, É.;~et~al.  Optical pump-rejection filter based on silicon sub-wavelength engineered photonic structures. \emph{Opt. Lett.} \textbf{2017}, \emph{42}, 1468. [\href{http://dx.doi.org/10.1364/OL.42.001468}{CrossRef}] [\href{http://www.ncbi.nlm.nih.gov/pubmed/28409775}{PubMed}]



\bibitem{savona} Minkov, M.; Savona, V.  A compact, integrated silicon device for the generation of spectrally filtered, pair-correlated photons.
\emph{J. Opt.} \textbf{2016}, \emph{18}, 054012. [\href{http://dx.doi.org/10.1088/2040-8978/18/5/054012}{CrossRef}]


\bibitem{spat} Wu, C.W.; Solntsev, A.S.; Neshev, D.N.; Sukhorukov, A.A.
Photon pair generation and pump filtering in nonlinear adiabatic waveguiding structures. \emph{Opt. Lett.} \textbf{2014}, \emph{39}, 953. [\href{http://dx.doi.org/10.1364/OL.39.000953}{CrossRef}] [\href{http://www.ncbi.nlm.nih.gov/pubmed/24562250}{PubMed}]

\bibitem{spat1} Solntsev, A.S.; Liu, T.; Boes, A.; Nguyen, T.G.; Wu, C.W.; Setzpfandt, F.; Mitchell, A.; Neshev, D.N.; Sukhorukov, A.A.
Towards on-chip photon-pair bell tests: Spatial pump filtering in a LiNbO3 adiabatic coupler.
\emph{Appl. Phys. Lett.} \textbf{2017}, \emph{111}, 261108. [\href{http://dx.doi.org/10.1063/1.5008445}{CrossRef}]

\bibitem{mitchell} Mitchell, M.; Lundeen, J.; Steinberg, A.   Super-resolving phase measurements with a multiphoton entangled state. \emph{Nature} \textbf{2004}, \emph{429}, 161–164. [\href{http://dx.doi.org/10.1038/nature02493}{CrossRef}]

\bibitem{dowling} Dowling, J.P.  Quantum optical metrology–the lowdown on high-N00N states. \emph{Contemp. Phys.} \textbf{2008}, \emph{49}, 125--143. [\href{http://dx.doi.org/10.1080/00107510802091298}{CrossRef}]

\bibitem{kowalewska} Kowalewska-Kudłaszyk, A.;  Leoński, W.;  Peřina, J., Jr. Generalized Bell states generation in a parametrically excited nonlinear coupler. \emph{Phys. Scr.} \textbf{2012}, \emph{2012}, 014016. [\href{http://dx.doi.org/10.1088/0031-8949/2012/T147/014016}{CrossRef}]

\bibitem{olsen} Olsen, M.K.  Spreading of entanglement and steering along small Bose-Hubbard chains. \emph{Phys. Rev. } \mbox{\textbf{2015}, \emph{92},~033627}. [\href{http://dx.doi.org/10.1103/PhysRevA.92.033627}{CrossRef}]

\bibitem{kalaga} Kalaga, J.K.; Leoński, W.; Szcz\c{e}śniak, R.  Quantum steering and entanglement in three-mode triangle Bose–Hubbard system. \emph{Quantum Inf. Process.} \textbf{2017}, \emph{16}, 265. [\href{http://dx.doi.org/10.1007/s11128-017-1717-5}{CrossRef}]

\bibitem{perina00} Pe{\v{r}}ina, J., Jr.; Pe{\v{r}}ina, J.  \textit{Quantum statistics of nonlinear optical couplers}; Progress in Optics; Wolf, E., Ed.;  Elsevier: Amsterdam, The Netherlands, 2000; p. 361.

\bibitem{korolkova97} Korolkova, N.; Pe{\v{r}}ina, J.  Quantum statistics and dynamics of Kerr nonlinear couplers.  \emph{Opt. Commun.} \textbf{1997},~\emph{136}, 135--149. [\href{http://dx.doi.org/10.1016/S0030-4018(96)00676-1}{CrossRef}]

\bibitem{korolkova97b}  Korolkova, N.; Pe{\v{r}}ina, J. Kerr nonlinear coupler with varying linear coupling coefficient. \emph{J. Mod. Opt.} \textbf{1997},~\emph{44}, 1525--1534. [\href{http://dx.doi.org/10.1080/09500349708230755}{CrossRef}]

\bibitem{thapliyal}Thapliyal, K.; Pathak, A.; Sen, B.; Pe{\v{r}}ina, J. Higher-order nonclassicalities in a codirectional nonlinear optical coupler: Quantum entanglement, squeezing, and antibunching.  \emph{Phys. Rev. A} \textbf{2014}, \emph{90}, 013808. [\href{http://dx.doi.org/10.1103/PhysRevA.90.013808}{CrossRef}]

\bibitem{mogilevtsev97} Mogilevtsev, D.; Korolkova, N.; Pe{\v{r}}ina, J.  Band-gap quantum coupler. \emph{J. Mod. Opt.} \textbf{1997}, \emph{44}, 1293--1307. [\href{http://dx.doi.org/10.1080/09500349708230738}{CrossRef}]

\bibitem{meany} Meany, T.; Gr{\"a}fe, M.; Heilmann, R.; Perez-Leija, A.; Gross, S.; Steel, M.J.; Withford, M.J.; Szameit, A.  Laser written circuits for quantum photonics. \emph{Laser Photonics Rev.} \textbf{2015}, \emph{9}, 363--384. [\href{http://dx.doi.org/10.1002/lpor.201500061}{CrossRef}]


\bibitem{mog2010} Mogilevtsev, D.; Shchesnovich, V.S. Single-photon generation by correlated loss in a three-core optical fiber. \emph{Opt. Lett.} \textbf{2010}, \emph{35}, 3375. [\href{http://dx.doi.org/10.1364/OL.35.003375}{CrossRef}]


\bibitem{natcom} Mukherjee, S.; Mogilevtsev, D.; Slepyan, G.Y.; Doherty, T.H.; Thomson, R.R.; Korolkova, N.  Dissipatively coupled waveguide networks for coherent diffusive photonics. \emph{Nat. Commun.} \textbf{2017}, \emph{8},  1909. [\href{http://dx.doi.org/10.1038/s41467-017-02048-4}{CrossRef}]


\bibitem{prap} Thornton, M.; Sakovich, A.; Mikhalychev, A.; Ferrer, J.D.; Hoz, P.d.; Korolkova, N.; Mogilevtsev, D.  Coherent Diffusive Photon Gun for Generating Nonclassical States.
\emph{Phys. Rev. Appl. }\textbf{2019}, \emph{12},  064051. [\href{http://dx.doi.org/10.1103/PhysRevApplied.12.064051}{CrossRef}]

\bibitem{valer} Shchesnovich, V.; Mogilevtsev, D. Three-site Bose-Hubbard model subject to atom losses: Boson-pair dissipation channel and failure of the mean-field approach. \emph{ Phys. Rev. A} \textbf{2010}, \emph{82}, 043621. [\href{http://dx.doi.org/10.1103/PhysRevA.82.043621}{CrossRef}]


\bibitem{nasu} Nasu, Y.; Kohtoku, M.; Hibino, Y. {Low-loss waveguides written
with a femtosecond laser for flexible interconnection in a planar light-wave circuit}. \emph{Opt. Lett}. \textbf{2005}, \emph{30}, 723. [\href{http://dx.doi.org/10.1364/OL.30.000723}{CrossRef}] [\href{http://www.ncbi.nlm.nih.gov/pubmed/15832918}{PubMed}]

\bibitem{grev} Gross, S.; Withford, M.J.
Ultrafast-laser-inscribed 3D integrated photonics: Challenges and emerging applications. \emph{Nanophotonics} \textbf{2015}, \emph{4}, 332. [\href{http://dx.doi.org/10.1515/nanoph-2015-0020}{CrossRef}]

\bibitem{pcf} Russell, P.S.J.  Photonic crystal fibers. \emph{Science} \textbf{2003}, \emph{299}, 358. [\href{http://dx.doi.org/10.1126/science.1079280}{CrossRef}] [\href{http://www.ncbi.nlm.nih.gov/pubmed/12532007}{PubMed}]

\bibitem{niels} Nielsen, M.D.; Jacobsen, C.; Mortensen, N.A.; Folkenberg, J.R.; Simonsen, H.R.   Low-loss photonic crystal fibers for transmission systems and their dispersion properties. \emph{Opt. Exp.} \textbf{2004}, \emph{12}, 1372. [\href{http://dx.doi.org/10.1364/OPEX.12.001372}{CrossRef}]


\bibitem{sil} Thomson, D.; Zilkie, A.; Bowers, J.E.; Komljenovic, T.; Reed, G.T.; Vivien, L.; Marris-Morini, D.; Cassan, E.; Virot, L.; Fedeli, J.-M.;~et~al. Roadmap on silicon photonics.  \emph{J. Opt.} \textbf{2016}, \emph{18}, 073003. [\href{http://dx.doi.org/10.1088/2040-8978/18/7/073003}{CrossRef}]

\bibitem{inp} Augustin, L.M.; Santos, R.; den Haan, E.; Kleijn, S.; Thijs, P.A.; Latkowski, S.; Zhao, D.; Yao, W.; Bolk, J.; Ambrosius, H.;~et~al.  InP-Based Generic Foundry Platform for Photonic Integrated Circuits.  \emph{IEEE J. Sel. Top.  Quantum Electron. }\textbf{2018},  \emph{24}, 10. [\href{http://dx.doi.org/10.1109/JSTQE.2017.2720967}{CrossRef}]



\bibitem{low}  Dong, P.; Qian, W.; Liao, S.; Liang, H.; Kung, C.; Feng, N.; Shafiiha, R.; Fong, J.; Feng, D.; Krishnamoorthy,~A.V.;~et~al.  Low loss shallow-ridge silicon waveguides. \emph{Opt. Exp.} \textbf{2010}, \emph{18}, 14474. [\href{http://dx.doi.org/10.1364/OE.18.014474}{CrossRef}]

\bibitem{sato} Sato, T.; Makino, S.; Ishizaka, Y.; Fujisawa, T.; Saitoh, K.
A rigorous definition of nonlinear parameter and effective area Aeff for photonic crystal optical waveguides. \emph{JOSA B} \textbf{2015},  \emph{32}, 1245. [\href{http://dx.doi.org/10.1364/JOSAB.32.001245}{CrossRef}]

\bibitem{agr} Agrawal, G.  \textit{Nonlinear Fiber Optics};  Elsevier Academic Press: Amsterdam, The Netherlands, 2013.


\bibitem{raman} Clark, A.S.; Collins, M.J.; Judge, A.C.; Magi, E.C.; Xiong, C.; Eggleton, B.J. Raman scattering effects on correlated photon-pair generation in chalcogenide.  \emph{Opt. Exp. }\textbf{2012}, \emph{20}, 16807. [\href{http://dx.doi.org/10.1364/OE.20.016807}{CrossRef}]

\bibitem{dezfouli} Dezfouli, M.K.; Dignam, M.M.; Steel, M.J.; Sipe, J.E.   Heisenberg treatment of pair generation in lossy coupled-cavity systems. \emph{Phys. Rev. A} \textbf{2014}, \emph{90}, 043832. [\href{http://dx.doi.org/10.1103/PhysRevA.90.043832}{CrossRef}]



\end{thebibliography}
\end{document}